\documentclass[a4paper,UKenglish,cleveref,autoref]{lipics-v2021}

\usepackage{amsmath,amssymb,amsfonts}
\usepackage{textcomp}
\usepackage{xcolor}
\usepackage{listings}
\usepackage{booktabs}
\usepackage{multirow}
\usepackage{graphicx}

\lstdefinelanguage{terraform}{
  keywords={resource, variable, output, provider, data, module, locals, terraform},
  sensitive=true,
  comment=[l]{\#},
  morecomment=[s]{/*}{*/},
  morestring=[b]",
}

\lstdefinestyle{diff}{
  basicstyle=\ttfamily\scriptsize,
  frame=single,
  framerule=0.4pt,
  breaklines=true,
  numbers=none,
  moredelim=**[is][\color{red!70!black}]{@-}{-@},
  moredelim=**[is][\color{green!50!black}]{@+}{+@},
}

\newcommand{\finding}[1]{\vspace{4pt}\noindent\fbox{\parbox{0.95\columnwidth}{\small\textbf{Finding:} #1}}\vspace{4pt}}

\title{Does Fixing Break Security? An Empirical Study of Security Degradation in Iterative LLM-Driven Infrastructure-as-Code Repair}
\titlerunning{Does Fixing Break Security?}

\author{Benjamin Agyekum}{Department of Electrical Engineering, Colorado State University, Fort Collins, CO, USA}{bagyekum@colostate.edu}{https://orcid.org/0009-0000-2937-5795}{}

\author{Fabio Santos}{Department of Computer Science, Colorado State University, Fort Collins, CO, USA}{fabio.deabreusantos@colostate.edu}{https://orcid.org/0000-0001-8069-3158}{}

\authorrunning{B. Agyekum and F. Santos}

\Copyright{Benjamin Agyekum and Fabio Santos}

\category{Technical Track Paper}


\relatedversiondetails[linktext={arXiv:2608.13404}]{Full Version}{https://arxiv.org/abs/2608.13404}

\EventEditors{Robert Feldt, Maria Paasivaara, Daniel Mendez, Stefan Wagner, and Marvin Mu\~{n}oz Bar\'{o}n}
\EventNoEds{5}
\EventLongTitle{20th International Symposium on Empirical Software Engineering and Measurement (ESEM 2026)}
\EventShortTitle{ESEM 2026}
\EventAcronym{ESEM}
\EventYear{2026}
\EventDate{October 8--9, 2026}
\EventLocation{Munich, Germany}
\EventLogo{}
\SeriesVolume{394}
\ArticleNo{49}

\begin{document}

\nolinenumbers

\maketitle

\begin{abstract}
\textbf{Background:} Iterative feedback loops have become the dominant paradigm for improving LLM-generated Infrastructure-as-Code (IaC): validators such as Checkov and \texttt{terraform validate} feed error signals back to the model for successive repair attempts. Prior work reports \emph{cumulative-best} metrics, which are monotonically non-decreasing by construction, so the raw per-iteration security trajectory has never been examined in the IaC domain.

\textbf{Aims:} We study \emph{security regression} (a previously-passing CIS Benchmark check that fails after a repair iteration) to determine whether, and how often, iterative LLM repair degrades security while fixing other issues.

\textbf{Method:} We analyze 5,968 scenario timelines from the IaC-Eval benchmark, each one scenario run through one configuration for up to 5 repair iterations. The 15 configurations comprise six model-specific RAG and nine model-aggregated non-RAG configurations, three temperatures each, and together they yield 4,440 iteration transitions with Checkov data on both sides. We track 30 individual CIS check IDs and classify regression root causes from code diffs, under two detection modes: \emph{standard} (inclusive) and \emph{strict} (exclusive check failures only).

\textbf{Results:} Under standard (inclusive) detection, 13.8\% of scenarios (24.8\% of transitions) exhibit at least one regression. Under strict detection, which counts only unambiguous, exclusive check failures, the rate falls to 3.3\% of scenarios (5.2\% of transitions). This gap indicates that most apparent regressions are multi-resource measurement artifacts rather than genuine exclusive failures. Resource restructuring (79.0\%) is the dominant root cause. Regression transitions show 2.6$\times$ more code churn (Cohen's $d=0.90$) and 4.9$\times$ higher strict-mode check volatility ($d=1.49$). Of standard-mode regressions, 36.6\% self-correct within an average of 1.2 iterations, and iteration~3 is the optimal stopping point.

\textbf{Conclusions:} Iterative IaC repair does introduce security regressions, but most \emph{apparent} regressions are multi-resource measurement artifacts. The conservative, defensible rate is approximately 3.3\% of scenarios. Our findings motivate security-aware feedback-loop design and provide actionable iteration-budget guidance.
\end{abstract}

\section{Introduction}\label{sec:intro}
Infrastructure-as-Code (IaC) has become the standard for managing cloud infrastructure, with Terraform serving as the leading declarative provisioning tool~\cite{hashicorp_terraform_main}. Ensuring security compliance of generated code remains a critical challenge, even after the rise of large language models (LLMs) that are increasingly adopted for automated IaC generation~\cite{palavalli2024using, xiang2025terrafault}. Studies report that, depending on prompting strategy and validation method, 12--65\% of LLM-generated code violates secure coding standards~\cite{das2024vulnerabilities_remediation}. IaC-specific evaluations find only 7\% of generated scripts are secure without explicit security guidance~\cite{milicevic2026security_7models}.

An increasingly explored approach to improve generated IaC is \emph{iterative feedback}: running static analysis tools (e.g., Checkov~\cite{checkov}, terraform validate) on the generated code and feeding error messages back to the LLM for successive repair attempts~\cite{palavalli2024using, xiang2025terrafault}. Prior work reports promising results from such feedback loops~\cite{palavalli2024using, xiang2025terrafault}. However, these studies uniformly report \emph{cumulative-best} metrics: the highest compliance achieved across all iterations. These are monotonically non-decreasing by construction and may mask important dynamics in the repair trajectory.

Recent studies on general-purpose code repair have raised alarms about \emph{security degradation}: the phenomenon where iterative LLM repair introduces new vulnerabilities while fixing existing issues. Shukla et al.~\cite{lam2025security_degradation} found a 37.6\% vulnerability increase after 5 iterations in C and Java, and Chen et al.~\cite{feng2026scaffold_cegis} demonstrated that 43.7\% of GPT-4o iteration chains contain more vulnerabilities than the baseline. Crucially, Chen et al. showed that Static Application Security Testing (SAST)-based gating can \emph{worsen} latent degradation, raising questions about whether Checkov-based feedback in IaC suffers similar effects.

Security regression is a known problem in traditional software evolution. Braz et al.~\cite{braz2022exploratory} found that regression vulnerabilities in Mozilla are introduced by bug fixes themselves, and security is rarely discussed while the fix is made. Felderer and Fourneret~\cite{felderer2015systematic} provide a taxonomy of security regression testing approaches. However, no study has investigated security regression specifically in the IaC domain with CIS compliance tracking. IaC differs fundamentally from general code: it is declarative rather than imperative, security checks map to specific configuration properties rather than data/control flow, and validators like Checkov operate at the resource level with enumerable check IDs. This resource-level structure is not a cosmetic difference: because one check can apply to several resources within a single configuration, IaC admits a class of \emph{multi-resource ambiguity} that general-code, CWE-based degradation studies cannot exhibit. As we show, this ambiguity accounts for the majority of apparent regressions. Studying regression in IaC therefore requires a detection methodology without direct analogue in prior general-code work: the \emph{standard}/\emph{strict} distinction we formalize in Section~\ref{sec:methodology}.

This paper presents the first large-scale empirical study of security regression in iterative LLM-driven IaC repair. We analyze 5,968 scenario timelines from the IaC-Eval benchmark~\cite{kon2024} across 15 configurations (two RAG models and three non-RAG prompting strategies, each run at three temperatures), tracking 30 individual Checkov check IDs mapped to CIS AWS Foundations Benchmark controls~\cite{cisbenchmark} across up to 5 repair iterations. Our study addresses four research questions:

\begin{itemize}
  \item \textbf{RQ1}: Does iterative LLM repair introduce security regressions in IaC?
  \item \textbf{RQ2}: What types of security controls are most vulnerable to regression?
  \item \textbf{RQ3}: How do prompting strategy, model, and temperature affect regression?
  \item \textbf{RQ4}: Is there a security-correctness trade-off in iterative IaC repair?
\end{itemize}

We make four contributions. This is the first large-scale empirical study of security regression in IaC repair, covering 5,968 scenario timelines, 4,440 transitions, and 30 check IDs across 15 configurations, based on single runs per configuration. From it we derive a taxonomy of regression root causes: in standard mode, resource restructuring accounts for 79.0\%, configuration drift 15.5\%, argument removal 3.6\%, and 1.9\% remain unclassified. We quantify the security-correctness trade-off, showing that regression transitions carry 2.6$\times$ more code churn and 4.9$\times$ higher strict-mode check volatility. Finally, we show that the standard and strict detection modes support qualitatively different conclusions, since the Mistral gap vanishes and RAG reverses between them, and we draw practical guidance from this: stop at iteration~3, expect 36.6\% of regressions to correct themselves, and prefer RAG where exclusive regressions matter most.

\section{Background and Motivation}\label{sec:background}

\textbf{CIS Benchmarks and Checkov:} The Center for Internet Security (CIS) AWS Foundations Benchmark v1.5.0~\cite{cisbenchmark} defines security controls for AWS infrastructure. Checkov~\cite{checkov} is a widely-adopted static analysis tool that maps these controls to concrete checks on Terraform configurations. Each Checkov check has a unique ID (e.g., \texttt{CKV\_AWS\_145} for S3 encryption) and can either PASS or FAIL for each resource in a Terraform plan. We map Checkov checks to six security categories: \emph{encryption}, \emph{access control}, \emph{logging}, \emph{networking}, \emph{data protection}, and \emph{other}.

\textbf{Iterative LLM-Based IaC Repair:}\label{sec:iterative-repair} The standard IaC feedback loop generates Terraform code from a natural-language prompt, validates it (\texttt{terraform validate} for syntax, Checkov for security), summarizes any errors back to the LLM for regeneration, and repeats for up to $K$ iterations. Each pair of consecutive iterations (e.g., 0$\to$1) constitutes an \emph{iteration transition}, the unit at which we measure regression. This paradigm is used by TerraFormer~\cite{jana2025terraformer} and Palavalli et al.~\cite{palavalli2024using}.

\begin{figure}[!ht]
  \centering
  \includegraphics[width=0.48\columnwidth, alt={Line chart comparing raw per-iteration and cumulative-best Checkov pass rates across iterations 0 to 5. The cumulative-best curve rises from 73 to 83 percent and never falls, while the raw curve peaks at 83.4 percent at iteration 4 and dips to 82.6 percent at iteration 5.}]{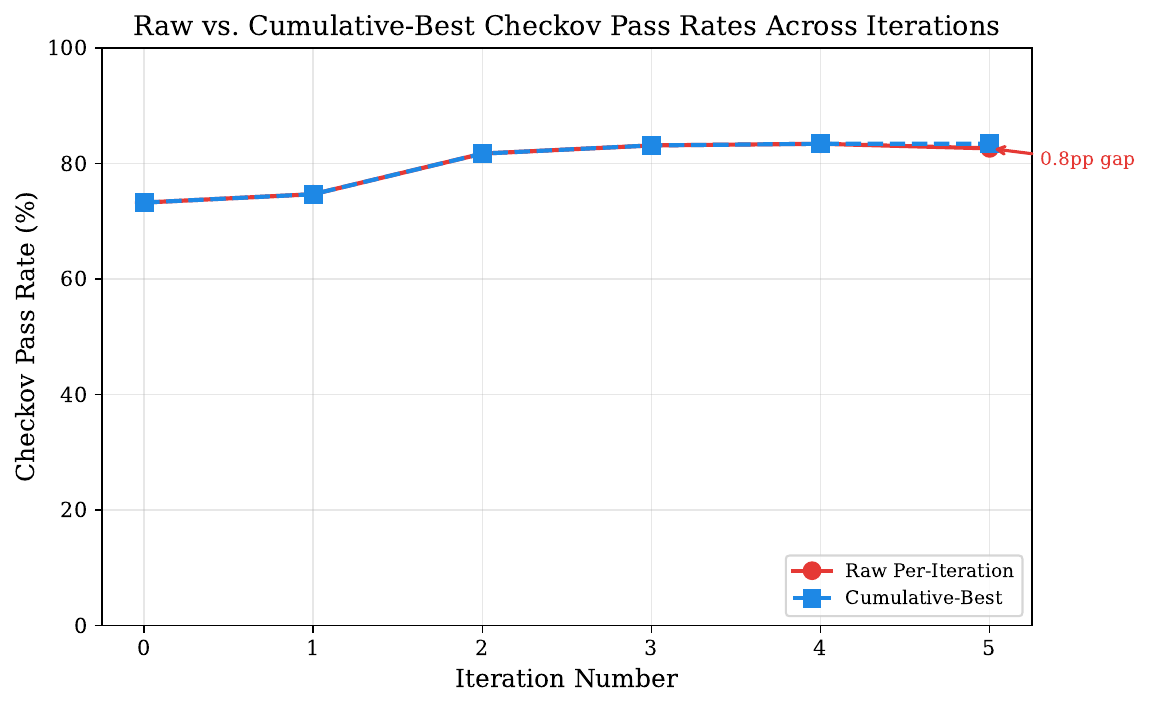}
  \caption{Raw per-iteration vs.\ cumulative-best Checkov pass rate (5,968 scenario timelines). The raw trajectory \emph{dips} at iteration~5 (82.6\% vs.\ 83.4\%), a regression cost masked by cumulative-best reporting.}
  \label{fig:motivating}
\end{figure}

\textbf{The Hidden Problem of Cumulative-Best Metrics:} Prior work on iterative IaC repair universally reports \emph{cumulative-best} metrics: the highest compliance rate achieved up to iteration $k$. These are monotonically non-decreasing by construction. The \emph{raw per-iteration trajectory}, however, may contain regressions. Figure~\ref{fig:motivating} illustrates this gap. Across 5,968 scenario timelines the cumulative-best CIS pass rate rises from 73\% to 83\%, yet the raw trajectory \emph{dips} at iteration~5 (82.6\% vs.\ the 83.4\% peak at iteration~4) as previously-passing checks fail.

\section{Related Work}\label{sec:related}

\textbf{Security of AI-Generated Code:} AI-generated code consistently shows security weaknesses: 29.5\% of Python Copilot snippets are vulnerable~\cite{fu2025security_copilot}, zero-shot security accuracy is only 37.44\%~\cite{safegenbench2025}, and 12--65\% of LLM code violates secure-coding standards~\cite{das2024vulnerabilities_remediation}. Sajadi et al.~\cite{he2025how_safe_patches} found LLMs introduce distinctive vulnerability patterns in automated patches, and Yan et al.~\cite{yan2025guiding} found that although over 98\% of GPT-4o vulnerability explanations are accurate, guided repair does not always eliminate the underlying issue. We instead track security \emph{changes across iterations} rather than static snapshots, in the IaC domain.

\textbf{RAG for Secure Code Generation:} RAG improves LLM code security by grounding generation in verified references. Sriram et al.~\cite{nong2026rag_multitool} combined RAG with multi-tool feedback (compiler, CodeQL, KLEE) and found retrieval context significantly reduces vulnerabilities, and RESCUE~\cite{dakhel2025rescue} retrieves secure coding patterns at generation time for substantial CWE reduction. Both show RAG's value for \emph{initial} security but not its stability across iterative repair. Our RAG pipeline grounds generation in CIS documentation and Terraform schema chunks. We find this yields more multi-resource outputs (raising standard-mode counts) but fewer exclusive failures (lowering strict-mode regressions).

\textbf{IaC-Specific Security and Compliance:} Recent IaC work spans generation, validation, and security. TerraFormer~\cite{jana2025terraformer} combines fine-tuning with policy-guided verifier feedback (+15.94\% on IaC-Eval), and Zhang et al.~\cite{zhang2025deployability} use DevOps simulation as feedback to improve deployment success. TerraFault~\cite{xiang2025terrafault} uses LLM agents to find bugs in IaC updates via plan-level diffs, complementary to our check-level analysis. Firouzi et al.~\cite{milicevic2026security_7models} found only 7\% of LLM-generated IaC secure without explicit guidance. GenSIaC~\cite{gensiac2025} and Diaz et al.~\cite{diaz2024} pursue security-aware generation and self-healing, while Palavalli et al.~\cite{palavalli2024using} showed feedback effectiveness decays exponentially to a plateau. To the best of our knowledge, we are the first to analyze the per-iteration security trajectory of IaC repair, tracking individual check regressions rather than aggregate compliance.

\textbf{Iterative Code Repair with LLMs:} Iterative LLM feedback is widely applied to code repair. Self-Refine~\cite{madaan2023selfrefine} showed $\sim$20\% gains from LLM self-critique. RepairAgent~\cite{bouzenia2025repairagent} reaches state-of-the-art bug repair on Defects4J via multi-tool orchestration, and LLMLOOP~\cite{ravi2025llmloop} improves functional correctness through generation--test feedback. Tang et al.~\cite{tang2024repair_exploration} model repair as an exploration--exploitation trade-off in which LLMs under-explore, and Cheng~\cite{cheng2026drv} proposes a Detect-Repair-Verify pipeline that evaluates repair side effects, finding repair can introduce new vulnerabilities. These works target functional correctness. None analyze whether iterative repair preserves security properties in IaC, which is our focus.

\textbf{Security Degradation in Iterative LLM Code Generation:} Security regression is well documented in traditional software evolution. Braz et al.~\cite{braz2022exploratory} studied 78 regression vulnerabilities at Mozilla and found they are introduced by bug fixes themselves.  Felderer and Fourneret~\cite{felderer2015systematic} give a taxonomy of security regression testing. For LLMs, Shukla et al.~\cite{lam2025security_degradation} found a 37.6\% vulnerability increase after five iterations of AI code generation, and Chen et al.~\cite{feng2026scaffold_cegis} showed that 43.7\% of GPT-4o iteration chains exceed baseline vulnerability counts and that SAST-based gating can \emph{worsen} latent degradation (12.5\%$\to$20.8\%).

Our study is best read as a conceptual replication and extension of Shukla et al.~\cite{lam2025security_degradation} in the IaC domain. We preserve the core design: iterative LLM repair with security measured after every iteration, making degradation visible inside the repair trajectory. We adapt the measurement apparatus to IaC, from imperative C and Java to declarative Terraform, and from multi-tool static analysis with manual review to Checkov's enumerable check IDs mapped to CIS controls. We extend the design with components absent from the original study: transition-level detection that judges each iteration against its immediate predecessor rather than cumulative vulnerability growth, dual standard/strict detection, a diff-based root-cause taxonomy, self-correction and oscillation analysis, retrieval-augmented configurations, and optimal-stopping analysis.


\section{Study Design}\label{sec:methodology}

\begin{figure}[t]
  \centering
  \includegraphics[width=0.8\linewidth, alt={Two-phase study overview. Phase 1, an iterative generation pipeline, runs left to right from the IaC-Eval dataset through the configuration space, LLM generation, Terraform code, and validation to iteration logs, with a feedback loop of up to five iterations returning from validation to generation. Phase 2, security regression analysis, runs from check tracking through regression detection, root cause classification, and statistical analysis to the findings for research questions 1 to 4.}]{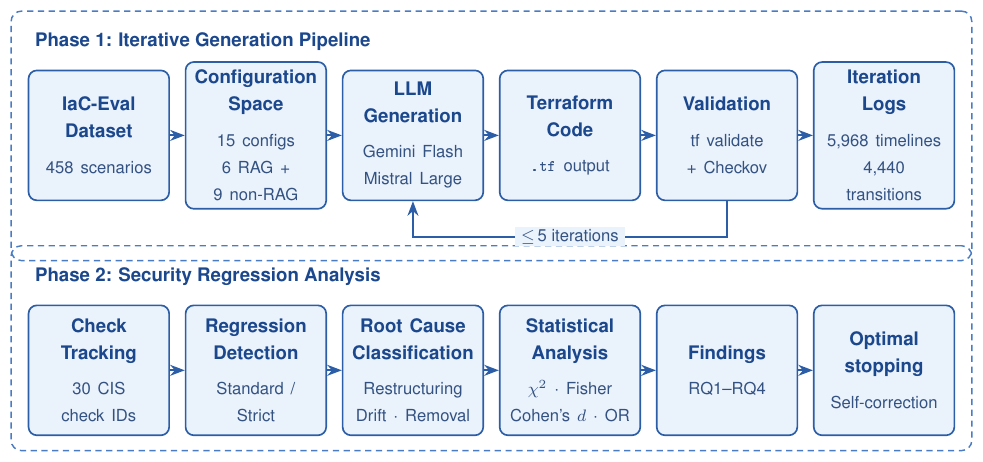}
    \caption{Study overview. Phase~1 generates Terraform through an iterative feedback loop (up to five iterations) across 15 configurations, yielding 5,968 timelines with 4,440 transitions. Phase~2 tracks 30 CIS check IDs to detect, classify, and quantify regressions under standard and strict modes.}
  \label{fig:architecture}
\end{figure}

Figure~\ref{fig:architecture} presents an overview of our two-phase methodology.

\subsection{Dataset and Configurations}\label{sec:dataset}

We used the IaC-Eval benchmark~\cite{kon2024} containing 458 Terraform generation scenarios spanning various AWS services. Each scenario specifies a natural-language infrastructure requirement to be translated into Terraform code.

We evaluated four prompting strategies at three temperatures, which, with RAG analyzed separately per model, give 15 analysis configurations.
\begin{itemize}
  \item \textbf{Prompting strategies} (4): Zero-Shot, Few-Shot, Chain-of-Thought (CoT), and RAG (Retrieval-Augmented Generation with CIS benchmark context). RAG is analyzed separately for each model, giving two groups, while each of the other three forms a single group, for five groups in total.
  \item \textbf{Temperatures} (3): 0.1, 0.4, and 0.7
\end{itemize}
These five groups at three temperatures give 15 configurations: six RAG (2 models $\times$ 3 temperatures), analyzed per model, and nine non-RAG (3 strategies $\times$ 3 temperatures), which are not. Both models were run with all four prompting strategies and neither was discarded. The difference is in what the experiment recorded. Every RAG run is stored together with the model that generated it, whereas the non-RAG runs are not, so an individual non-RAG run cannot be traced back to Gemini or Mistral. Where a scenario was executed more than once under the same non-RAG strategy and temperature, we keep the most recent execution, so each non-RAG timeline is one run from one model rather than an average of the two. For the 36\% of non-RAG timelines whose model can still be determined, the split is 66\% Gemini to 34\% Mistral (893 and 461 of 1,354), so these groups lean toward Gemini rather than being balanced. We analyze each as a single group whose model composition is estimated rather than known.

Running one scenario under one configuration produces a \emph{timeline}: the ordered sequence of code versions from the initial generation (iteration~0) through up to five repair iterations. A timeline ends early once the code passes all checks or the iteration budget is exhausted, so timelines range from a single generation to six versions.

\textbf{Configuration rationale:} The four strategies span the guidance spectrum of prior IaC-generation work: Zero-Shot provides only the task and a security-oriented system prompt. Few-Shot adds three input--output examples. Chain-of-Thought adds an explicit reasoning trace to essentially the same examples. RAG moves to per-query retrieved context from CIS documentation and Terraform schemas. The first three steps each add a single ingredient, permitting descriptive comparison. RAG changes several things at once (its own system-prompt template, no static examples, retrieved context), so we compare it as a paradigm rather than one more increment. 

Gemini 2.0 Flash and Mistral Large Latest are production-oriented models from independent vendors, both publicly available at data-collection start (December 2024). Temperatures 0.1, 0.4, and 0.7 cover the lower and middle part of the 0--1 range explored in code-generation evaluations, where low temperatures suit the single-sample setting we use, and higher ones mainly help when many samples are drawn per task~\cite{chen2021codex}. The negligible effect we observe (Cram\'er's $V = 0.049$) is consistent with reports that sampling temperature has little influence on task accuracy~\cite{renze2024temperature}.

We excluded agent-based and self-refinement approaches because they replace the validator-in-the-loop paradigm whose per-iteration behavior we measure (Section~\ref{sec:iterative-repair}).

We ran each configuration up to five feedback iterations per scenario, validating with Terraform v1.6.0 (\texttt{terraform validate}) and Checkov v3.2.392. Versions were locked at data collection start, December 2024, for reproducibility. Every iteration was recorded with its full Terraform code and individual check results. We limited the feedback loop to five iterations, following prior IaC feedback work that found repair effectiveness plateaus after roughly five iterations~\cite{palavalli2024using}. Our regression analysis reconstructs timelines from these per-iteration records, which is where the model attribution described above is lost. We therefore analyzed the six RAG configurations at the model level, Gemini and Mistral, and, for each non-RAG strategy and temperature, analyzed one timeline per scenario. This precludes a model-controlled comparison outside the RAG setting (see Section~\ref{sec:discussion}).

The three non-RAG strategies shared a single system prompt that instructed the model to generate CIS-compliant Terraform, combining general directives (least-privilege IAM, encryption at rest, logging, restricted network access) with resource-specific guidance (e.g., configuring S3 public-access controls through a separate \texttt{aws\_s3\_bucket\_public\_access\_block} resource). Few-Shot appended three input--output pairs that map a natural-language request to complete HCL (e.g., ``Create an AWS RDS instance \ldots{} with randomly generated id and password''). Chain-of-Thought appended essentially the same three example scenarios, each preceded by a reasoning trace that enumerated the required resources, filled in their attributes, and then wired the resources together. RAG used a separately maintained system-prompt template covering the same core security directives, into which the retrieved CIS-control and resource-schema chunks are injected at query time. The complete templates are included in the replication package~\cite{replication-package}. Both models were run with all four strategies. As noted above, only the RAG logs record which model produced each iteration.

\subsection{Regression Detection}\label{sec:detection}

We define a \emph{security regression} as follows:

\begin{definition}[Security Regression]
Given consecutive iterations $i$ and $i{+}1$ of a scenario where both have Checkov results, a regression occurs for check $c$ if $c \in \text{PASSED}(i)$ and $c \in \text{FAILED}(i{+}1)$.
\end{definition}

Transitions are formed only between consecutive iterations with \emph{distinct} indices. Where a scenario has two records sharing an iteration index, that pair is excluded: it reflects a repeated attempt at the same step rather than a repair step. We employed two detection modes. \textbf{Standard}: any check ID that is in the passed set at iteration $i$ and in the failed set at iteration $i{+}1$ counts as a regression, regardless of whether it also appears in the passed set (multi-resource scenarios). \textbf{Strict}: A check must be \emph{exclusively} passed (not in failed) in iteration $i$ and \emph{exclusively} failed (not in passed) in iteration $i{+}1$. This excludes ambiguous multi-resource cases.
Strict detection trades sensitivity for unambiguity: it can miss genuine degradations that affect only a subset of a configuration's resources. 
For example, suppose two S3 buckets pass CKV\_AWS\_145 (S3 encryption) at iteration~$i$; at $i{+}1$ the LLM restructures the code, and one bucket now fails the check. \emph{Standard} mode counts this as a regression (the check is in both the passed and failed sets). \emph{Strict} mode does not (CKV\_AWS\_145 is not exclusively failed: it still passes for one bucket). We return to this trade-off in Section~\ref{sec:std-vs-strict}.

\subsection{Metrics}



Five metrics quantify regression prevalence, the size of the code changes that accompany it, and how often the loop recovers. The \emph{scenario regression rate} is the fraction of scenarios with $\geq$1 regression event, and the \emph{transition regression rate} the fraction of transitions with $\geq$1 regression. \emph{Code churn} covers both lines changed and the text similarity ratio between consecutive iterations, where 0 is completely different and 1 identical. \emph{Check volatility} is $V = |C_{\text{new}}| + |C_{\text{removed}}| + |C_{\text{flipped}}|$, where $C_{\text{new}}$ are checks appearing for the first time at iteration $i{+}1$, $C_{\text{removed}}$ are checks present at iteration $i$ but absent at $i{+}1$, and $C_{\text{flipped}}$ are checks changing status, either pass$\to$fail or fail$\to$pass. The \emph{self-correction rate} is the fraction of regressed checks that return to passing later in the same scenario timeline, reported in Section~\ref{sec:deepdive}.

For RQ4 we additionally examined two derived transition classes. A \emph{post-syntax-fix transition} is one whose prior iteration was syntactically invalid and whose current iteration is valid for the first time, isolating the effect of syntax repair on security. A \emph{syntax-struggled timeline} is a scenario whose iteration sequence contained at least one invalid-syntax iteration, as opposed to a \emph{clean timeline} that was syntactically valid throughout. We compared regression rates between these two groups.

The self-correction rate, root-cause taxonomy, and temporal patterns are reported as deeper analyses in Section~\ref{sec:deepdive}.

\subsection{Root Cause Classification}

For each regression transition, we classify the root cause from the code diff using four mutually exclusive categories, in order of priority. A transition is \emph{resource restructuring} if the number of Terraform resource blocks changes by $\geq$2, or if resource types appear or disappear between iterations. The threshold of 2 separates structural rewrites from single-resource edits, where a $\pm$1 block change is common and would over-trigger the category. Failing that, it is \emph{argument removal} if security-relevant arguments were deleted: the classifier scans removed diff lines for patterns in five groups, covering encryption (\texttt{encrypt*}, \texttt{kms\_key}, \texttt{storage\_encrypted}, \texttt{ssl\_policy}), access control (\texttt{policy}, \texttt{iam\_role}, \texttt{acl}, \texttt{block\_public}), logging (\texttt{logging}, \texttt{flow\_log}, \texttt{cloudwatch}, \texttt{trail}), networking (\texttt{security\_group}, \texttt{cidr\_block}, \texttt{ingress}, \texttt{waf}), and versioning or backup (\texttt{versioning}, \texttt{backup}, \texttt{deletion\_protection}). The full regular-expression set is in the replication package~\cite{replication-package}. Failing both, it is \emph{configuration drift} if text similarity between iterations exceeds 0.85, meaning small changes flipped a check's status. The 0.85 cutoff separates near-identical edits, such as a single property change, from substantial rewrites, and we treat it as a heuristic (Section~\ref{sec:discussion}). Anything left, such as a diff that failed to parse, is \emph{unclassified}.

\subsection{Statistical Methods}

We used chi-squared tests for categorical associations, Mann-Whitney $U$ tests for continuous variables (code churn, volatility), Fisher's exact test for 2$\times$2 tables, and Wilson score intervals for confidence intervals. Effect sizes are reported as Cram\'er's $V$ (categorical), Cohen's $d$ (continuous, used for interpretability despite non-parametric testing), and odds ratios (binary). All tests used $\alpha = 0.05$. We apply Bonferroni family-wise error rate control within each research question. RQ2 corrects $k=6$ post-hoc category comparisons ($\alpha_{\text{adj}} = 0.0083$). RQ3 corrects $k=4$ omnibus configuration-factor tests (model, strategy, RAG vs.\ non-RAG, temperature) and RQ4 $k=4$ independent tests, both at $\alpha_{\text{adj}} = 0.0125$. All reported effects remain significant after correction. Detailed test specifications are in the replication package~\cite{replication-package}.

\subsection{Scale}

Table~\ref{tab:dataset} summarizes our dataset. Of a theoretical maximum of $458 \times 15 = 6{,}870$ scenario-configuration pairs, 902 (13\%) are absent due to API rate-limit errors, provider timeouts, or scenarios that produced no parseable output. Applying the deduplication and same-iteration rules (Sections~\ref{sec:dataset} and~\ref{sec:detection}) leaves 5,968 scenario timelines and 4,440 iteration transitions with Checkov data on both sides.

\begin{table}[t]
\caption{Dataset Summary}
\label{tab:dataset}
\centering\small
\begin{tabular}{lr}
\toprule
\textbf{Metric} & \textbf{Value} \\
\midrule
IaC-Eval scenarios & 458 \\
Configurations analyzed & 15 \\
Max iterations per scenario & 5 \\
Total iteration log files parsed & 36,736 \\
Unique scenario timelines & 5,968 \\
Transitions with Checkov data & 4,440 \\
Individual Checkov check IDs tracked & 30 \\
CIS security categories & 6 \\
\bottomrule
\end{tabular}
\end{table}

\section{Results}\label{sec:results}
The four research questions build on one another: RQ1 measures how often regressions occur, RQ2 where they concentrate, RQ3 which configurations amplify them, and RQ4 what code-level behavior accompanies them. Section~\ref{sec:deepdive} then examines causes, timing, and stopping policy.
\subsection{RQ1: Does Iterative LLM Repair Introduce Security Regressions?}

In the standard analysis, of the 5,968 scenario timelines, 823 (13.8\%, 95\% CI: [12.9\%, 14.7\%]) exhibit at least one security regression event. Across the 4,440 transitions with Checkov data on both sides, 1,103 (24.8\%, CI: [23.6\%, 26.1\%]) contain at least one regression, producing 2,639 total regression events.

Under strict detection (excluding ambiguous multi-resource cases), 194 scenarios (3.3\%, CI: [2.8\%, 3.7\%]) exhibit regression, with 282 total events across 231 transitions (5.2\% of transitions, CI: [4.6\%, 5.9\%]). Scenarios fall 4.2$\times$, from 823 to 194. Transitions fall comparably, 1,103 to 231. Approximately 76\% of standard-mode-flagged scenarios are not flagged in strict mode, consistent with multi-resource ambiguity rather than exclusive check failures. We use the scenario-level 76\% as the headline artifact fraction.


We measure the fraction of scenarios with $\geq$1 regression event (13.8\% standard, 3.3\% strict), while Shukla et al.~\cite{lam2025security_degradation} measure cumulative vulnerability growth (37.6\% after 5 iterations in C and Java). Metric differences limit direct comparison. The lower IaC rates may reflect its declarative nature and structured validation, but a definitive comparison requires metric alignment across domains.

\finding{13.8\% of IaC scenario timelines experience security regression in iterative repair, lower than general code (37.6\%) but still substantial. Strict analysis shows 3.3\%, indicating most regressions involve multi-resource ambiguity.}

\subsection{RQ2: What Types of Security Controls Are Most Vulnerable?}

RQ1 established how often regressions occur. RQ2 asks where they land. Table~\ref{tab:rq2_categories} shows the regression distribution across security categories. Regression events are highly concentrated rather than spread evenly across the six categories: a chi-squared goodness-of-fit test against a uniform reference rejects an even split ($\chi^2 = 1445$, $p < 0.001$, Cram\'er's $V = 0.33$, medium-to-large effect). We report this test only to quantify the skew (uniformity is not a theoretically motivated null) and use the per-category breakdown below to identify its drivers.
\begin{table}[t]
\caption{Regression events by security category (Standard / Strict)}
\label{tab:rq2_categories}
\centering\small
\begin{tabular}{lrrrr}
\toprule
\textbf{Category} & \multicolumn{2}{c}{\textbf{Standard}} & \multicolumn{2}{c}{\textbf{Strict}} \\
\cmidrule(lr){2-3} \cmidrule(lr){4-5}
 & Count & \% & Count & \% \\
\midrule
Access control & 1,017 & 38.5 & 26 & 9.2 \\
Data protection & 684 & 25.9 & 7 & 2.5 \\
Encryption & 357 & 13.5 & 61 & 21.6 \\
Networking & 375 & 14.2 & 120 & 42.6 \\
Logging & 152 & 5.8 & 38 & 13.5 \\
Other & 54 & 2.0 & 30 & 10.6 \\
\midrule
\textbf{Total} & \textbf{2,639} & & \textbf{282} & \\
\bottomrule
\end{tabular}
\end{table}

In standard mode, \emph{access control} dominates (38.5\%), driven by three IAM-related checks: CKV\_AWS\_356 (298 events), CKV\_AWS\_111 (297), and CKV\_AWS\_109 (291). These checks evaluate IAM policy conditions, wildcard permissions, and role trust boundaries, properties frequently modified during code restructuring.

Strikingly, the category ranking \emph{reverses} in strict mode: \emph{networking} (42.6\%) and \emph{encryption} (21.6\%) dominate, while access control drops to 9.2\%. This reversal indicates that access control regressions are predominantly multi-resource ambiguity, while networking and encryption represent genuine exclusive check failures. The mechanism is structural: IAM scenarios typically generate multiple resources (e.g., \texttt{aws\_iam\_role} and \texttt{aws\_iam\_policy}) where a check like CKV\_AWS\_356 may pass for one resource but fail for another, creating standard-mode regressions without exclusive failure. In contrast, networking checks (e.g., VPC endpoints) and encryption checks (e.g., KMS key rotation) typically map one-to-one to resources, making multi-resource ambiguity rare.

\textbf{Operational impact.} Not all regressing checks carry equal risk. The three most frequently regressing checks all govern IAM privilege boundaries, among the highest-impact misconfigurations operationally: CKV\_AWS\_356, CKV\_AWS\_111, and CKV\_AWS\_109 (886 events, 33.6\% of all standard-mode regressions). Aggregating by category, the high-impact categories (access control, encryption, networking) account for 1,749 events (66.3\%), while advisory logging checks account for only 5.8\%. This coarse lens indicates that regressions concentrate in the controls practitioners care about most, not in low-stakes advisory checks. A full severity- or exploitability-weighted analysis is future work.

\finding{Access control dominates standard regressions (38.5\%) but drops to 9.2\% in strict mode. Networking and encryption represent the most genuine regression risks, and the most frequent regressing checks (886 events, 33.6\%) all govern IAM privilege.}

\subsection{RQ3: How Do Configuration Factors Affect Regression?}
Knowing how often (RQ1) and where (RQ2) regressions strike, we next ask which configurations amplify them, examining four factors: model, prompting strategy, retrieval augmentation, and temperature. One caveat applies throughout. The logs record the generation model only for RAG runs, so non-RAG results aggregate over an unidentified model and may partly reflect model behavior rather than the named factor. We flag this in each affected result and return to it in Section~\ref{sec:discussion}.


In standard mode, the model effect is dramatic: Gemini's scenario regression rates range from 1.9\% to 4.7\% across temperatures, while Mistral ranges from 32.6\% to 41.2\%. Mistral scenarios are over 17 times more likely to regress (OR $= 17.29$, $p < 0.001$). However, this gap \textbf{vanishes entirely} in strict mode: neither RAG+Gemini nor RAG+Mistral produces a single strict-mode scenario regression (0 of 1,297 and 0 of 929 respectively), indicating that Mistral's elevated standard-mode rate stems from multi-resource ambiguity rather than exclusive check failures. RAG+Mistral transitions also exhibit substantially larger code changes than RAG+Gemini (median 32 vs.\ 8 lines changed), consistent with Mistral regenerating whole resource blocks and thus producing more multi-resource configurations.


Regression rates differ significantly across the five strategy/model groups. A 5$\times$2 test of the five strategy/model groups in Table~\ref{tab:rq3_config} against regressed vs.\ non-regressed scenario counts gives $\chi^2 = 596$, dof~$=4$, $p < 0.001$, Cram\'er's $V = 0.32$ (medium effect). Among non-RAG strategies, Chain-of-Thought achieves the lowest regression rates (8.2--9.6\%), followed by Few-Shot (11.6--12.9\%) and Zero-Shot (11.6--15.5\%), which would suggest that more structured prompting reduces regression risk. That reading does not survive the model composition of these groups. Among the non-RAG timelines whose model can be determined (Section~\ref{sec:methodology}), Chain-of-Thought is 79\% Gemini, Few-Shot 73\%, and Zero-Shot only 46\%, the same order as the regression ranking. Because Gemini regresses far less often than Mistral under RAG (1.9--4.7\% vs.\ 32.6--41.2\%), a ranking of this shape is what differing model mixes alone would produce. We therefore report the ordering as descriptive of the non-RAG configurations as deployed and draw no causal conclusion about prompting strategy.
\begin{table}[t]
\caption{Scenario regression rates by configuration and temperature. $N$~= scenarios analyzed. Reg~= scenarios with $\geq$1 regression. RAG rows are reported per model. Non-RAG rows are not split by model (Section~\ref{sec:methodology}).}
\label{tab:rq3_config}
\centering\scriptsize
\setlength{\tabcolsep}{4pt}
\begin{tabular}{lrrrrrrrrr}
\toprule
 & \multicolumn{3}{c}{$t=0.1$} & \multicolumn{3}{c}{$t=0.4$} & \multicolumn{3}{c}{$t=0.7$} \\
\cmidrule(lr){2-4}\cmidrule(lr){5-7}\cmidrule(lr){8-10}
\textbf{Configuration} & $N$ & Reg & Rate & $N$ & Reg & Rate & $N$ & Reg & Rate \\
\midrule
RAG + Gemini & 413 & 8 & 1.9\% & 447 & 21 & 4.7\% & 437 & 15 & 3.4\% \\
RAG + Mistral & 138 & 45 & 32.6\% & 439 & 161 & 36.7\% & 352 & 145 & 41.2\% \\
Zero-Shot & 354 & 55 & 15.5\% & 458 & 53 & 11.6\% & 458 & 62 & 13.5\% \\
Few-Shot & 419 & 54 & 12.9\% & 336 & 39 & 11.6\% & 458 & 53 & 11.6\% \\
Chain-of-Thought & 453 & 37 & 8.2\% & 369 & 33 & 8.9\% & 437 & 42 & 9.6\% \\
\bottomrule
\end{tabular}
\end{table}


The RAG effect \emph{reverses} between modes. In standard mode, RAG has a \emph{higher} regression rate: a 2$\times$2 test of RAG vs.\ non-RAG scenarios against regressed-vs-non-regressed counts gives OR $= 1.67$, $p < 0.001$. In strict mode, however, RAG records \emph{zero} scenario regressions, versus 4--6\% for the non-RAG strategies. This means RAG produces \emph{fewer exclusive regressions} but generates more multi-resource code structures that inflate the standard count (see Section~\ref{sec:discussion} for the retrieval mechanism explanation). Note that this comparison is not controlled for model: the non-RAG timelines are not attributable to a specific model in our logs, whereas RAG includes both Gemini and Mistral.


Temperature has a statistically significant but practically negligible effect (a 3$\times$2 contingency test over the three temperatures: $\chi^2 = 14.3$, $p < 0.001$, Cram\'er's $V = 0.049$). Higher temperatures show slightly elevated regression rates, but the effect size is minimal.

\finding{The model gap (Mistral 17$\times$ worse than Gemini in standard mode) vanishes entirely in strict mode (neither RAG model produces a single strict regression), revealing it as a multi-resource artifact. RAG reverses from worse to better between standard and strict analysis.}

\subsection{RQ4: The Security-Correctness Trade-off}

The remaining question is mechanism: RQ4 tests whether fixing correctness comes at a security cost, from four angles: the code churn and check volatility of regressing transitions, the regression rate of transitions that immediately follow a syntax fix, and the regression rate of timelines that struggled with syntax.


We compare code churn between all Checkov-bearing transitions that contain a regression and those that do not, pooled across every strategy, model, and temperature. Transitions with regressions show significantly more code modification than those without: the average number of lines changed between consecutive iterations is 140.1 vs.\ 53.8, a 2.6$\times$ difference. The text similarity ratio between iterations also drops from 0.847 to 0.791, indicating that regression transitions involve more extensive rewrites. Mann-Whitney $U$ tests confirm significance ($p < 0.001$, Cohen's $d = 0.90$ for lines changed, large effect).


Check volatility measures the total number of check-level changes between consecutive iterations (Section~\ref{sec:methodology}). It is strongly elevated in regression transitions. In standard mode, average volatility is 5.17 with regression vs.\ 2.09 without, a 2.5$\times$ difference. In strict mode it is even stronger: 11.55 vs.\ 2.38 (4.9$\times$), with Cohen's $d = 1.49$, a very large effect and the single strongest signal in our study, making check volatility the strongest indicator of security regression.


In transitions immediately following a syntax fix, the prior iteration had invalid syntax, and the current iteration first achieves validity. These show a 21\% higher regression rate than transitions between two already-valid iterations (28.1\% vs.\ 23.3\%, OR $= 1.29$, $p < 0.001$). The two groups partition the 4,440-transition universe: 1,449 and 2,991 transitions. Their regression counts, 407 and 696, sum to 1,103, the count underlying the 24.8\% aggregate rate. This effect is not significant in strict mode after Bonferroni correction ($p = 0.045$, above the adjusted $\alpha = 0.0125$), suggesting that post-fix regressions are predominantly multi-resource ambiguity introduced when the LLM restructures code to achieve validity.


Finally, we analyze the syntax struggle paradox. Counter-intuitively, scenarios that \emph{always} had valid syntax (``clean'' timelines) show a \emph{higher} regression rate than those that struggled with syntax errors: clean timelines regress in 20.8\% of cases versus 12.1\% for syntax-struggled timelines. A 2$\times$2 test (clean vs.\ syntax-struggled timelines $\times$ scenarios with $\geq$1 regression vs.\ none) gives OR $= 1.90$ ($p < 0.001$). We discuss two candidate mechanisms for this counter-intuitive result in Section~\ref{sec:discussion}.

\finding{Larger code changes strongly predict regression (2.6$\times$ more churn, $d=0.90$). Check volatility is the strongest predictor ($d=1.49$ strict). Syntax-fixing transitions carry 21\% higher regression risk.}

\section{Deep Dive Analysis}\label{sec:deepdive}

This section reports analyses that extend the four research questions rather than introducing new ones: the root-cause taxonomy and temporal patterns deepen RQ1 and RQ2 by characterizing \emph{how} and \emph{when} regressions arise, while the optimal-stopping analysis operationalizes the RQ4 security-correctness trade-off into a concrete iteration budget.

\subsection{Root Cause Taxonomy}

We classified the root cause of every regression transition (1,103 standard and 231 strict) from the code diff between consecutive iterations, using resource-block changes, text similarity, and deleted security arguments (Section~\ref{sec:methodology}). Table~\ref{tab:rootcauses} shows the distribution.
\begin{table}[t]
\caption{Root cause classification of regression transitions}
\label{tab:rootcauses}
\centering\small
\begin{tabular}{lrrrr}
\toprule
\textbf{Root Cause} & \multicolumn{2}{c}{\textbf{Standard}} & \multicolumn{2}{c}{\textbf{Strict}} \\
\cmidrule(lr){2-3} \cmidrule(lr){4-5}
 & Count & \% & Count & \% \\
\midrule
Resource restructuring & 871 & 79.0 & 158 & 68.4 \\
Configuration drift & 171 & 15.5 & 46 & 19.9 \\
Argument removal & 40 & 3.6 & 19 & 8.2 \\
Unclassified & 21 & 1.9 & 8 & 3.5 \\
\bottomrule
\end{tabular}
\end{table}

\textbf{Resource restructuring} (79.0\%) is the dominant cause. The LLM adds, removes, or renames resource blocks while fixing validation errors, and the restructured code loses security configurations from the prior iteration. This is especially common in access control, reflecting how tightly IAM policies are coupled to resource structure.

\textbf{Configuration drift} (15.5\%) involves small modifications that flip a check's status, typically changing a single property value or adding/removing an attribute. It most often affects networking and encryption checks, where small changes to security-group rules or KMS key references can fail specific checks.

\textbf{Argument removal} (3.6\% standard, 8.2\% strict) represents explicit deletion of security-relevant Terraform arguments. Its proportion more than doubles in strict mode, indicating these are genuine security regressions rather than ambiguous multi-resource artifacts.

\textbf{Temporal Patterns.} Beyond \emph{which} checks regress, we examine \emph{when} regressions occur across the iteration sequence, whether they are subsequently undone, and how often checks flip repeatedly.

\textbf{When Do Regressions Occur?}

\begin{figure}[t]
  \centering
  \includegraphics[width=0.50\columnwidth, alt={Bar chart of regression event counts by iteration transition. Counts are highest at the earliest transitions, 769 events at 0 to 1, 711 at 1 to 2, and 616 at 2 to 3, then decline steadily through the later transitions as code stabilizes.}]{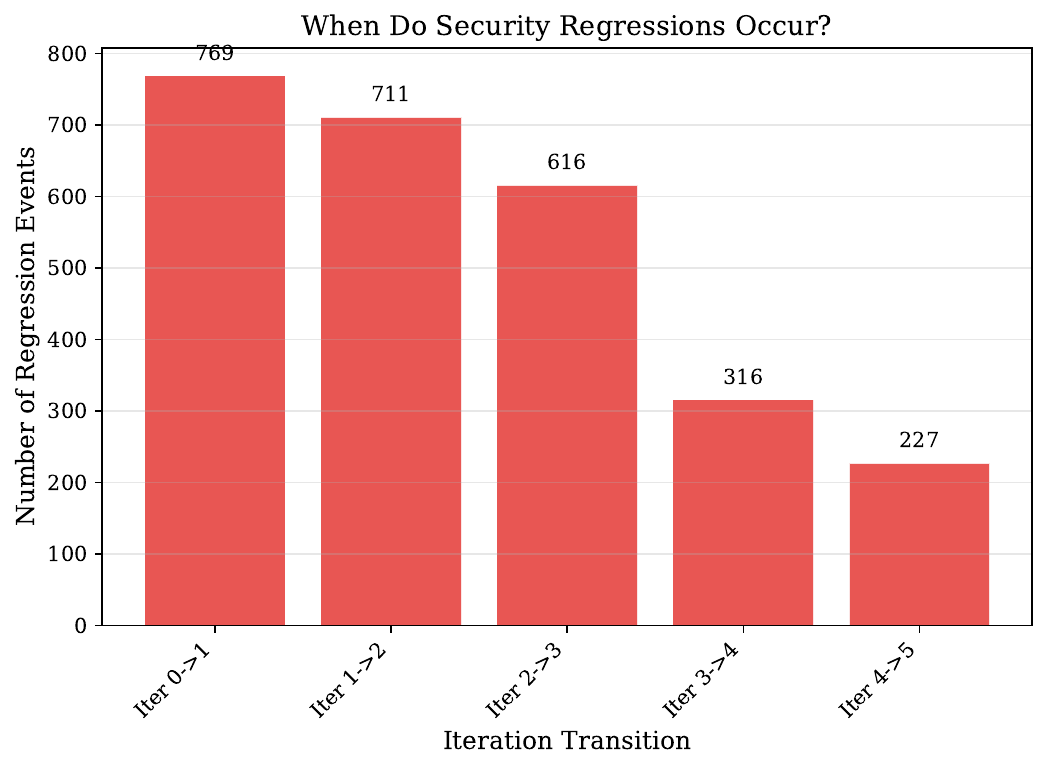}
\caption{Regression events by iteration transition: counts peak in early-to-middle transitions, then decline as code stabilizes.}
  \label{fig:temporal}
\end{figure}

Figure~\ref{fig:temporal} shows the distribution of regression events across iteration transitions. The highest counts occur early (0$\to$1: 769 events, 1$\to$2: 711, 2$\to$3: 616), reflecting the larger number of scenarios still iterating. Normalized by per-slot transition counts, however, the per-transition \emph{rate} peaks at 2$\to$3 (29.5\%) and 1$\to$2 (26.6\%), while 0$\to$1 is lower (22.8\%), consistent with the 24.8\% aggregate rate. One plausible explanation, unconfirmed without per-transition modification analysis, is that early iterations make conservative fixes while middle iterations restructure more aggressively. Late iterations (4$\to$5: 20.8\%) show lower rates, possibly reflecting code stabilization.

\textbf{Self-Correction.}
Of 2,639 standard-mode regressions, 967 (36.6\%) self-correct in a subsequent iteration, meaning the regressed check returns to passing status later in the same scenario's repair timeline. On average, self-correction takes 1.2 iterations after the regression occurs (median: exactly 1), with 80.4\% correcting within a single step (i.e., at the next repair iteration). This indicates that the feedback loop often catches and reverses its own mistakes. Self-correction is at least as common in strict mode (44.0\%, mean 1.2 iterations). We report the standard-mode 36.6\% as the headline figure because its far larger event pool (2,639 vs.\ 282 events) gives a more reliable estimate.

\textbf{Check Oscillation.} 
28.5\% of all scenarios (1,698 of 5,968) exhibit \emph{oscillating} checks, i.e., checks that pass, then fail, then pass again (or vice versa) across iterations. The rate is identical under strict detection, as oscillation is a timeline property independent of the multi-resource threshold. The three most oscillation-prone checks are all IAM-related: CKV\_AWS\_356 (925 timelines), CKV\_AWS\_111 (922), and CKV\_AWS\_109 (832), suggesting the LLM lacks stable strategies for complex IAM configurations and cycles between implementations satisfying different subsets of checks.

\textbf{Optimal Stopping Point.} 
Table~\ref{tab:stopping} combines average pass rate and cumulative regression data to identify the optimal stopping point. The ``N'' column counts timelines with Checkov data at each iteration: it rises from iteration~0 to~1 (2,649 $\to$ 3,151) as initially invalid scenarios become checkable, then falls as scenarios reach full compliance or exhaust their budget.

\begin{table}[t]
\caption{Compliance vs.\ regression trade-off by iteration (standard mode). N varies as scenarios terminate upon full compliance or budget exhaustion. Average cumulative regressions may decrease when high-regression scenarios terminate earlier, leaving a lower-regression survivor pool.}
\label{tab:stopping}
\centering\small
\begin{tabular}{crrrr}
\toprule
\textbf{Iter} & \textbf{N} & \textbf{Avg Pass} & \textbf{Avg Cum.\ Reg} & \textbf{\% w/ Reg} \\
\midrule
0 & 2,649 & 73.2\% & 0.00 & 0.0\% \\
1 & 3,151 & 74.7\% & 0.28 & 13.0\% \\
2 & 2,726 & 81.7\% & 0.49 & 17.8\% \\
3 & 1,878 & 83.1\% & 0.68 & 20.8\% \\
4 & 1,393 & 83.4\% & 0.64 & 18.8\% \\
5 & 1,029 & 82.6\% & 0.94 & 23.3\% \\
\bottomrule
\end{tabular}
\end{table}

Iteration~3 offers the best trade-off. We anchor this recommendation on the pass-rate trajectory, which is the more reliable of the two signals: the pass rate reaches 83.1\% at iteration~3, within 0.3pp of the 83.4\% maximum at iteration~4, and then \emph{decreases} to 82.6\% at iteration~5. If we define marginal gains as $<$0.5pp (with ${\sim}$30 tracked checks, 0.5pp $\approx$ 0.15 checks), the iteration~3$\to$4 gain falls below this threshold, so iteration~3 captures essentially all attainable improvement.

The average cumulative-regression column tells a consistent but weaker story, and we treat it as secondary because it is \emph{not monotone}: it rises 0.00$\to$0.68 through iteration~3, dips to 0.64 at iteration~4, then rises to 0.94 at iteration~5. The dip is a survivor-bias artifact, since high-regression scenarios disproportionately terminate by iteration~3 and leave a lower-regression survivor pool, so the column alone cannot serve as a stopping criterion. It does show that regressions keep accumulating past iteration~3. This matches Palavalli et al.'s~\cite{palavalli2024using} finding of exponential feedback decay and suggests that adaptive stopping policies (e.g., halt when $\Delta$pass\_rate $<$ 0.5pp for two consecutive iterations) could optimize the compliance-regression trade-off per scenario.

\section{Discussion}\label{sec:discussion}
Our results tell a single story. Iterative repair does break previously-passing checks (RQ1), concentrated in IAM, networking, and encryption controls (RQ2). The damage tracks how much code is rewritten far more than any configuration knob (RQ3--RQ4). Much of what standard detection flags is multi-resource bookkeeping rather than exclusive failure. The implications below follow from this reading.
The following design hypotheses for IaC tooling are motivated by our observations. None has been implemented or evaluated in this study. We offer them as directions for future tooling rather than validated solutions.

\textbf{Security anchoring.} The dominant root cause, resource restructuring (79.0\%), suggests that LLMs should be constrained to make \emph{minimal modifications} during repair rather than regenerating entire resource blocks. A ``security anchor'' mechanism could lock passing checks and only modify failing ones.

\textbf{Check-aware feedback.} Current feedback loops report all errors without distinguishing new failures from pre-existing ones. Explicitly flagging regressions (``Warning: CKV\_AWS\_145 was passing but now fails'') could prevent the LLM from unknowingly degrading security.

\textbf{Iteration budgets.} Our finding that iteration~3 is the optimal stopping point supports a conservative iteration budget. Tools should not blindly iterate to convergence. Three iterations capture most of the compliance improvement while keeping regression risk manageable.

\textbf{Choosing a configuration.} Our results support cost- and risk-differentiated guidance rather than a single recommendation. Where retrieval infrastructure is unavailable or too costly, Chain-of-Thought showed the lowest non-RAG regression rates (8.2--9.6\%) at no infrastructure cost beyond a longer prompt. That ranking tracks the model mix of the non-RAG groups rather than the strategies themselves (Section~\ref{sec:results}), so it is descriptive rather than causal. Where stability of individual security properties is paramount, RAG is preferable, since it produced zero strict-mode regressions. Its price is maintaining a retrieval index and triaging the standard-mode alerts that its multi-resource output generates. In either case the iteration budget above applies. Check volatility is cheap to compute online and is the strongest regression signal in our data ($d = 1.49$, strict mode). Halting or flagging a repair loop when volatility spikes offers regression protection at negligible cost. A single regression event, by contrast, is a premature halting signal: 36.6\% of regressions self-correct within the loop.

\textbf{Standard vs.\ Strict: Two Valid Perspectives.}\label{sec:std-vs-strict}
The dramatic differences between standard and strict analysis (13.8\% vs.\ 3.3\% scenario rate, Mistral 17$\times$ worse in standard mode vs.\ zero RAG regressions in strict) reflect a fundamental tension in how regression is defined for multi-resource IaC. Standard mode captures the practitioner view: any degradation of a check for \emph{any} resource counts, while strict mode captures the analyst view: only unambiguous, exclusive regressions count. Strict mode's exclusivity requirement carries a security cost: a check that fails for several resources while still passing for one is invisible to strict detection, so a transition that degrades most, but not all, instances of a control is not counted. Strict mode is thus a measurement-validity device, not an operational security criterion: it isolates regressions that cannot be explained by multi-resource bookkeeping, at the deliberate price of under-counting genuine partial degradations. We therefore recommend reporting both modes in distinct roles: the strict figures anchor the conservative point estimate we defend as our headline result, while the standard figures give the inclusive outer bound an operator should act on: production monitoring should triage standard-mode alerts precisely because strict detection can miss partial degradations, whereas research comparison and cross-study benchmarking should use strict mode.

\textbf{How Real Is the Phenomenon?}\label{sec:how-real}
A skeptical reading must confront the gap between our standard and strict figures. The standard-mode rates (13.8\% of scenarios, 24.8\% of transitions) are an \emph{inclusive upper bound}: roughly three quarters (76\% of standard-mode regression scenarios, a comparable 79.1\% reduction at the transition level) are multi-resource bookkeeping artifacts rather than exclusive check failures (Section~\ref{sec:results}). The conservative, defensible claim is the strict rate: approximately 3.3\% of scenarios (5.2\% of transitions) exhibit an unambiguous, exclusive regression.

Three caveats compound around even the strict rate. We ran each configuration once rather than repeating the experiment, so run-to-run variance is unquantified. The root-cause taxonomy is automated diff-classifier output, never validated against human judgment. Some timelines may mix iterations from separate executions. Together these mean the strict 3.3\% should be read as an order-of-magnitude estimate rather than a precise point value. What the data \emph{does} support robustly is the qualitative claim: iterative IaC repair sometimes degrades a previously-passing security check, the effect is not negligible, and cumulative-best reporting hides it entirely.

\textbf{Self-Correction and Oscillation.}
The 36.6\% self-correction rate implies that single-iteration regression is an imperfect halting signal: roughly one-third of regressions are transient, resolved by the same feedback mechanism that caused them. The 28.5\% oscillation rate, however, shows many corrections are themselves unstable. A check may regress, recover, and regress again, so not all self-corrections are stable recoveries. The three most oscillation-prone checks (CKV\_AWS\_356, CKV\_AWS\_111, CKV\_AWS\_109) are all IAM policy checks, suggesting the LLM cycles between implementations satisfying different subsets of IAM constraints. Stabilizing them needs more structured retrieval context or constrained generation.

\textbf{RAG Reversal and Temperature Effects.}
The RAG reversal effect (RAG appears \emph{worse} in standard mode but \emph{better} in strict mode) reveals that RAG generates more multi-resource configurations while producing fewer exclusive regressions. We attribute this to retrieval: a query mentioning ``VPC with logging'' retrieves schema chunks for \texttt{aws\_vpc}, \texttt{aws\_flow\_log}, and related log-storage resources. That context encourages comprehensive multi-resource solutions, increasing the opportunity for check ambiguity in standard mode. However, the same contextual grounding provides stability for individual security properties, explaining why RAG achieves fewer exclusive regressions in strict mode. The negligible temperature effect (Cram\'er's $V = 0.049$) likewise suggests regression is driven by structural generation patterns rather than sampling randomness.

\textbf{Why Clean Timelines Regress More.}
A counter-intuitive RQ4 result is that scenarios with consistently valid syntax (``clean'' timelines) regress \emph{more} often than those that struggled with syntax errors (OR~$= 1.90$). We identify two non-mutually-exclusive explanations. First, an \emph{opportunity effect}: clean timelines have more consecutive valid-to-valid transitions, providing more opportunities for Checkov-level regressions to occur, since only syntactically valid code undergoes security evaluation. Second, a \emph{feedback purity effect}: when the LLM receives only security feedback (no syntax errors), it aggressively restructures already-valid code in pursuit of compliance, inadvertently breaking passing checks. Disentangling these mechanisms would require an ablation study varying feedback types, which we leave to future work.

\textbf{What Generalizes Beyond Terraform and Checkov?}
Which findings are specific to Terraform and Checkov, and which plausibly characterize iterative LLM repair at large? Multi-resource ambiguity, and with it the standard/strict gap and the mode-dependent reversals of RQ3, arises because one check can apply to several resources in a configuration. We expect it to transfer to other IaC ecosystems (CloudFormation, Ansible, tfsec, Terrascan), which likewise evaluate checks per resource, but not to general-purpose code, where findings are not multiplied across resource instances. By contrast, three observations are candidates for domain-general behavior: large rewrites are strongly associated with regressions, restructuring rather than minimal editing dominates, and repair gains diminish across iterations. The first two are consistent with the degradation Shukla et al.~\cite{lam2025security_degradation} and Chen et al.~\cite{feng2026scaffold_cegis} observe in general-purpose code, and the third is replicated within IaC by Palavalli et al.'s feedback plateau~\cite{palavalli2024using}. Whether self-correction and oscillation recur outside IaC is unexamined. These are hypotheses: confirming them requires replications that vary domain and validator while holding the repair loop fixed.

From a human-factors perspective, the contrast is instructive. Braz et al.~\cite{braz2022exploratory} found that developers focus on the complexity of the bug at hand and work under community pressure to deliver, while security goes undiscussed during the fix. The mechanism we observe differs: LLM regressions arise within a single repair step, when the model rewrites structure wholesale rather than editing minimally. The blind spot, however, is shared. Just as security went unexamined during Mozilla's bug fixes, current feedback loops never tell the model which checks its previous iteration was already passing.

\textbf{Threats to Validity}

\textbf{Internal:} Our root-cause classification relies on automated diff analysis: it was not validated against human labels, may misclassify boundary cases (e.g., resource renaming), and its thresholds ($>$0.85 similarity for drift, $\geq$2 resource-count change for restructuring) were chosen conservatively rather than empirically tuned. Alternatives could shift the distributions. We mitigate this with the dual standard/strict framework, but manual validation of a stratified sample remains future work. The per-iteration logs record the generation model for RAG runs but not non-RAG runs, precluding a model-controlled RAG-vs-non-RAG comparison and the isolation of retrieval-augmentation effects from model behavior. We partially mitigate this by reporting RAG+Gemini and RAG+Mistral separately (Table~\ref{tab:rq3_config}). Where a scenario ran more than once we keep the later execution and drop same-iteration transitions, yet consecutive iterations in such a timeline may still come from different attempts. The 902 absent scenario-configuration pairs (13\%) stem from API errors, timeouts, and unparseable outputs, and are unlikely to bias our estimates. We use single runs per configuration for computational reasons, since repeated trials (e.g., $n=3$) would require roughly 1.3M additional API calls, and although the negligible temperature effect ($V=0.049$) suggests stability, point estimates may vary $\pm$2--5pp. Finally, we treat transitions as independent for Mann-Whitney tests despite possible within-timeline autocorrelation, our iteration-3 stopping point rests on a subjective $<$0.5pp threshold, and the post-hoc root-cause taxonomy and exploratory subgroup analyses were not pre-registered.

\textbf{External:} Our results are specific to Terraform and Checkov. Other IaC tools (Ansible, CloudFormation) and validators (tfsec, Terrascan) may exhibit different regression patterns. Our dataset (IaC-Eval v1.0, 458 scenarios, accessed December 2024) covers AWS only. We ran 24 experiments in total, all four strategies on both models at three temperatures. Only the six RAG experiments record which model produced each run, so the eighteen non-RAG experiments aggregate into nine model-aggregated configurations, giving the 15 we analyze (Section~\ref{sec:methodology}). Attributing all 24 would enable cleaner causal inference. Our findings may also partly reflect the repair capability of the two studied models. More recent or reasoning-oriented models may restructure less aggressively or preserve passing checks more reliably. Replication with newer models is needed to separate paradigm-level from model-level effects.
\textbf{Construct:} Checkov provides a proxy for security. Real security assessment requires deployment-time evaluation (\texttt{terraform plan}, sandbox testing). We use static validation rather than deployment for scalability (5,968 deployments would incur substantial cost), reproducibility (outcomes vary by account state), and safety (some scenarios create production resources). Some Checkov checks may be overly strict or context-dependent. Our study also compares RAG holistically without isolating individual components (CIS docs vs.\ Terraform schema vs.\ retrieval depth). Future ablation studies should test these variants separately.

\section{Conclusion}\label{sec:conclusion}

We presented the first large-scale empirical study of security regression in iterative LLM-driven IaC repair, analyzing 5,968 scenario timelines across 15 configurations with up to 5 repair iterations each. Six findings emerge. (1)~\textbf{Regression is real:} 13.8\% of scenarios regress in standard mode, 3.3\% in strict mode. (2)~\textbf{Detection mode changes conclusions:} access control dominates standard mode while networking and encryption are the genuine strict-mode risks, the 17$\times$ Mistral-vs-Gemini gap vanishes, and RAG reverses from worst to best. Multi-resource ambiguity drives standard-mode differences. (3)~\textbf{Resource restructuring} is the dominant root cause (79.0\%), ahead of configuration drift (15.5\%). (4)~\textbf{Code churn and check volatility are indicators of regressions}, with strict-mode volatility the strongest signal ($d=1.49$). (5)~\textbf{Self-correction is common but unstable:} 36.6\% of standard-mode regressions self-correct, yet 28.5\% of scenarios oscillate. (6)~\textbf{Iteration~3 is the optimal stopping point}, balancing an 83.1\% pass rate against manageable regression risk. Together, these findings provide actionable guidance for designing security-aware IaC feedback loops.
\section*{Data Availability}
Our replication package is openly available~\cite{replication-package}. It contains the analysis scripts, the derived results that reproduce every table, figure, and statistic reported here, the complete prompt templates, a \texttt{README} documenting the pipeline, and a slimmed subset of the raw per-iteration logs so the pipeline can be exercised end to end.
\bibliography{reference}

\end{document}